\documentclass[reprint,aps,twocolumn,epsfig,eqsecnum]{revtex4-1}
\usepackage{graphicx}
\newcommand{\assign}{:=}

\newcommand\be{\begin{eqnarray}}
\newcommand\ee{\end{eqnarray}}
\newcommand\ba{\begin{array}}
\newcommand\ea{\end{array}}

\newcommand{\ket}[1]{|#1\rangle} %ket
\newcommand{\bra}[1]{\langle#1|} %bra

\def\r{\rangle}
\def\l{\langle}

\def\tr{{\rm Tr}}

\def\cE{{\cal E}}
\def\cF{{\cal F}}
\def\cG{{\cal G}}
\def\cH{{\cal H}}
\def\cI{{\cal I}}

\def\cK{{\cal K}}
\def\cL{{\cal L}}

\def\cS{{\cal S}}

\begin{document}

%%%%%%%%%%%%%%%%%%%%%%%%%%%%%%%%%%%%%%%%%%%%%%% BEGIN TITLE %%%%%%%%%%%%%
\title{Direct estimation of decoherence rates}
\author{Vladim\'\i r Bu\v zek$^{1}$, Peter Rap\v can$^{1}$, Jochen Rau$^{2,3}$, M\'ario Ziman$^{1,4}$}
\address{
$^{1}$Research Center for Quantum Information, Slovak Academy of Sciences,
D\'ubravsk\'a cesta 9, 845 11 Bratislava, Slovakia \\
$^{2}$ Institut f\"{u}r Angewandte Physik,
Technische Universit\"{a}t Darmstadt, 
Hochschulstr. 4a, 
64289 Darmstadt, Germany \\
$^{3}$ Institut f\"{u}r Theoretische Physik, Johann Wolfgang Goethe-Universit\"{a}t, Max-von-Laue-Str. 1, 60438 Frankfurt am Main, Germany \\
$^{4}$ Institute of Theoretical Physics, ETH Z\"urich, Switzerland
}
%\ead{ziman@savba.sk}
\begin{abstract}
The decoherence rate is a nonlinear channel parameter that
 describes quantitatively the decay of the off-diagonal elements 
of a density operator in the decoherence basis. 
We address the question of how to experimentally access such a nonlinear 
parameter directly without the need of complete process tomography. 
In particular, we design a simple experiment working with two copies 
of the channel, in which the registered mean value of a two-valued 
measurement directly determines the value of the average decoherence 
rate. No prior knowledge of the decoherence basis is required.
\end{abstract}
\maketitle
%%%%%%%%%%%%%%%%%%%%%%%%%%%%%%%%%%%%%%%%%%%%%%% END TITLE %%%%%%%%%%%%%%%%%%

%%%%%%%%%%%%%%%%%%%%%%%%%%%%%%%%%%%%%%%%%%%%%%%%%%%%%%%%%%%%%%%%%%%%%%%%%%%%%
\section{Introduction}
%%%%%%%%%%%%%%%%%%%%%%%%%%%%%%%%%%%%%%%%%%%%%%%%%%%%%%%%%%%%%%%%%%%%%%%%%%%%%
Complete quantum tomography is a very 
complex task which rapidly becomes computationally intractable as the number 
of subsystems composing the system under investigation increases (see e.g., Ref. \cite{nielsen}). More specifically, if we assume that $d$ is the 
dimension of each of the subsystems, then a system composed of $N$ such subsystems
has dimension $D=d^N$. The goal of state (or channel) tomography is then to determine $D^2-1$ or $D^2(D^2-1)$ real parameters, 
respectively. That is, the number of parameters grows exponentially
with the number $N$ of quantum subsystems that comprise the complete system. As a consequence, already 
for few-qubit systems the tomography of associated quantum devices requires  an almost insurmountable  
experimental and computational effort.

This enormous effort is the reason why physicists put forth tasks that are more modest  than complete tomography. 
For instance, quite  often the system under consideration is not completely unknown; rather, some vital  prior information is available.
If so, the estimation problem can become tractable. 
This idea has led to the development 
of biased estimation schemes \cite{gross_csense,toth,plenio_mpsest,rau:evidence}
that work efficiently (i.e., with resource requirements scaling only polynomially with system size) for states within a certain class of ``expected states''. 
Other research lines do without estimating all system parameters and properties and instead focus on 
identifying  parameters that are experimentally feasible, i.e.,  
quantities that can be "easily" measured \cite{audenaert+plenio,wunderlich:incomplete}. 
Here we focus on one such parameter, the (average) decoherence rate, and discuss
the minimal (informationally incomplete) resources needed for its estimation.

The phenomenon of decoherence is often recognized as the main  
obstacle to the experimental implementation of  quantum information technologies. In its essence, it converts 
a quantum superposition (with associated amplitudes) into an incoherent mixture
(with associated probabilities). Mathematically, a loss of quantum coherence 
is reflected by a decrease of the off-diagonal terms (irrespective of their initialization) of the system's
density operator   
in the so-called decoherence basis, while the diagonal elements 
remain preserved. In the present paper we will assume that the estimated 
processes are of this form, and we will focus on possible methods of 
estimation
of their parameters, especially 
of the so-called decoherence rates, defined as fractions of absolute 
values of the final and initial values of off-diagonal elements. These parameters
are not linear, which opens interesting questions on their direct
experimental accessibility. Although in this paper we will consider 
only a very special case of nonlinear parameters, our discussion and 
findings are aiming to develop a general mathematical framework dealing 
with such type of estimation problems.

In the next section we will briefly describe the relevant properties
of decoherence channels. In Section III we will focus on an experimental setup to access directly 
the decoherence rate (being
a nonlinear parameter) for qubit channels. We will generalize this 
consideration to the case of $d$-dimensional quantum systems in Section IV. 
The achieved results will be applied
in Section V to direct estimation of the (inverse) decoherence rate of a 
double-commutator  master equation. In the final Section VI we will 
briefly summarize our results and discuss a possible experimental realization.

%%%%%%%%%%%%%%%%%%%%%%%%%%%%%%%%%%%%%%%%%%%%%%%%%%%%%%%%%%%%%%%%%%%%%%%%%%%%%
\section{Decoherence channels}
%%%%%%%%%%%%%%%%%%%%%%%%%%%%%%%%%%%%%%%%%%%%%%%%%%%%%%%%%%%%%%%%%%%%%%%%%%%%%
Let us denote the Hilbert space of the considered system by $\cH$. By
$\cL(\cH)$ we denote the set of linear operators on $\cH$, with
$\cS(\cH)$ being the subset of density operators 
($\tr{\varrho}=1,\varrho\geq 0$). States are identified with
elements of $\cS(\cH)$, and (deterministic) quantum processes are described by 
quantum channels, i.e., by completely positive trace-preserving 
linear maps $\cE:\cL(\cH)\to\cL(\cH)$. These properties guarantee that quantum 
channels map quantum states into quantum states, i.e., 
$\cE(\cS(\cH))\subset\cS(\cH)$.
A distinguished role within the set of quantum channels is played by
unitary channels for which $\cE(\varrho)=U\varrho U^\dagger$, where
$UU^\dagger=U^\dagger U=I$. It is quite common to identify
\emph{decoherence channels} with non-unitary ones, although for most of the 
channels there is no orthonormal basis that is invariant under the channel's 
action. If such a basis exists we say that the (non-unitary) channel is a 
\emph{pure decoherence channel}. 

Elementary properties of pure decoherence channels were studied 
in Refs. \cite{buscemi2005,ziman2005}. Pure decoherence channels are 
parametrized by a choice of the decoherence basis and a choice of (complex) 
inverse decoherence rates, the latter independently for each (mutually conjugated) pair of
off-diagonal terms. In particular, under the action of a pure decoherence 
channel, elements of the density operator expressed in the decoherence basis
undergo the transformation
\be
\varrho_{jk}\mapsto\varrho_{jk}^\prime=\omega_{jk}\varrho_{jk}\,,
\label{eq:DensMatTransformation}
\ee
where $|\omega_{jk}|$ are the inverse decoherence rates ($ j\ne k$) and, in general, $\omega_{jk}=\overline{\omega}_{kj}$, $|\omega_{jk}|\leq 1$ and $\omega_{jj}=1$; thus the transformation is characterized 
by $D(D-1)$ real parameters. Moreover, we need to specify additional 
$D(D-1)$ parameters in order to determine the  (unordered) decoherence basis. 
Altogether, we see that a pure decoherence channel is characterized 
by $2D(D-1)$ real parameters, which is significantly less than 
for a general channel but  
still exponential in the number of the system's constituents, even if the decoherence
basis is known. Our goal is to design an 
experiment that would allow us to determine  the value of just a single parameter
which characterizes the ``average'' decoherence. Namely,
we will investigate measurement of the quantity $\overline{\lambda^2}:=
\frac{2}{d(d-1)}\sum_{j<k}|\omega_{jk}|^2$, which for qubits
reduces to estimation of the inverse decoherence rate $\lambda$. 
Simultaneously, we require that the experiment provides 
as little information as possible about any other parameters. 
In an ideal case the inverse decoherence rate $\lambda$ will be measured as the mean 
value (or its function) of a specific two-valued measurement. If this
happens to be the case, we say the parameter is measured directly.

%%%%%%%%%%%%%%%%%%%%%%%%%%%%%%%%%%%%%%%%%%%%%%%%%%%%%%%%%%%%%%%%%%%%%%%%%%%%%
\section{Qubit case}
%%%%%%%%%%%%%%%%%%%%%%%%%%%%%%%%%%%%%%%%%%%%%%%%%%%%%%%%%%%%%%%%%%%%%%%%%%%%%
The most general qubit pure decoherence channel can be written in
the following form \cite{ziman2005}
\be
\varrho\to\varrho^\prime=\cE(\varrho)=p \varrho+(1-p)U\varrho U^\dagger,
\ee
where $U$ is an arbitrary unitary operator. The eigenbasis of $U$
determines the decoherence basis. Suppose $U=e^{ia}\ket{0}\bra{0}+
e^{ib}\ket{1}\bra{1}$. Then $\omega_{01}=p+(1-p)e^{i(a-b)}$, and the 
(unique) qubit inverse decoherence rate reads 
$\lambda=|\omega_{01}|=\sqrt{\omega_{01}\omega_{10}}$. 
Our goal is to propose an ``optimal'' measurement that would allow us to determine  this parameter directly,
while measuring as little redundant other information as possible.
In order to achieve our goal we have  to specify an initial probe state $\omega$ 
and a two-valued observable (outcomes associated with effects $E,I-E$) 
such that 
\be
\label{eq:OrigProblem}
\lambda=f\left(\tr{(\cE_\lambda\otimes\cI)[\omega]E}\right)\,,
\ee 
where $f$ is a function on the interval $[0,1]$. According to the 
Choi-Jamiolkowski isomorphism, linear maps $\cF$ from $\cL(\cH)$ to
$\cL(\cK)$ are in one-to-one correspondence with linear operators $F$
defined on $\cH\otimes\cK$. In particular, 
$F=(\cF\otimes\cI)[\Omega_+]$ with $\Omega_+=
\sum_{j,k=1}^d\ket{j\otimes j}\bra{k\otimes k}\in\cH\otimes\cH$ 
being an unnormalized version of the corresponding maximally entangled state. 
The complete positivity of channels translates into positivity 
of the operators. Writing $\omega=(\cI\otimes\cF)[\Omega_+]$,
$\Omega_\lambda=(\cE_\lambda\otimes\cI)[\Omega_+]$ 
and $F=(\cI\otimes\cF^*)[E]$, Eq. (\ref{eq:OrigProblem}) can be rewritten
in the form
\be
\label{eq:problem}
\lambda=f\left(\tr{\Omega_\lambda F}\right)\,,
\ee
where $F$ is an element of a so-called process POVM (PPOVM) (introduced
in Refs. \cite{ziman_ppovm,dariano_testers}) which fully captures
the adjustable degrees of freedom (choice of initial probe 
state and final measurement) in the experiment.

%%%%%%%%%%%%%%%%%%%%%%%%%%%%%%%%%%%%%%%%%%%%%%%%%%%%%%%%%%%%%%%%%%%%%%%%%%%%%
\subsection{Known decoherence basis}
%%%%%%%%%%%%%%%%%%%%%%%%%%%%%%%%%%%%%%%%%%%%%%%%%%%%%%%%%%%%%%%%%%%%%%%%%%%%%
Let us suppose that the decoherence basis is known to be the one in which
the operator $\Omega_+$ is defined. Then its Choi-Jamiolkowski
representation reads
\be
\Omega_\lambda=\ket{00}\bra{00}+\ket{11}\bra{11}+
\omega_{01}\ket{00}\bra{11}+\omega_{10}\ket{11}\bra{00}\,.
\ee
It follows from Eq. (\ref{eq:problem}) that no single Hermitian
operator $Q$ allows for a determination of $\lambda=
\sqrt{\omega_{01}\omega_{10}}=\tr{[\Omega_\lambda Q]}$,
simply because this parameter is not linear. However, 
there are several ways to determine the value 
of $\lambda$ by combining expectation values
of more than one Hermitian operator.  Since $\lambda=
|\bra{11}\Omega_\lambda\ket{00}|=:|\tr{\Omega_\lambda T}|$, the decoherence parameter $\lambda$
can be accessed ``directly''  by measuring the mean value of a non-Hermitian
operator $T=\ket{00}\bra{11}$. In practice, this requires to realize an experiment according to the following recipe:\\
1) Prepare a maximally entangled bipartite state 
and send one of its subsystems through the decoherence
channel. \\
2) Define observables $X:=\{X_+,X_-,X_o\}$ 
and $Y:=\{Y_+,Y_-,Y_o\}$, where 
$X_\pm:=\frac{1}{2}(\ket{00}\pm \ket{11})(\bra{00}\pm\bra{11})$, 
$X_o:=I-X_+-X_-$  and
$Y_\pm:=\frac{1}{2}(\ket{00}\pm i\ket{11})(\bra{00}\pm i\bra{11})$, 
$Y_o:=I-Y_+-Y_-$. \\
3) Then $T=\frac{1}{2}[(X_+-X_-)+i(Y_+-Y_-)]$; hence
in this case, the evaluation of $\lambda$ requires the experimental 
specification of two independent probabilities, because 
for the given initial states the outcomes 
$X_o=Y_o=\ket{01}\bra{01}+\ket{10}\bra{10}$ cannot appear.

The following examples will demonstrate that if one uses the observed
probabilities in a nonlinear way 
(i.e., by taking a multivariate function $f=f(X_1,\dots,X_n)$), 
there are even simpler experiments 
that reveal the value of $\lambda$. The $X_1,\dots,X_n$ are linear 
operators whose expectation values $\tr{\Omega_\lambda X_j}$ 
are experimentally measured.
For example, 
initialize the input state to be $\ket{+}=(\ket{0}+\ket{1})/\sqrt{2}$.
The measurement of $\sigma_x=\ket{0}\bra{1}+\ket{1}\bra{0}$ 
and $\sigma_y=-i(\ket{0}\bra{1}-\ket{1}\bra{0})$ results in expectation values
\be
x&=&\tr{\left[\cE_\lambda[\ket{+}\bra{+}]\sigma_x\right]}\,,\\
y&=&\tr{\left[\cE_\lambda[\ket{+}\bra{+}]\sigma_y\right]}\,,
\ee
respectively. One can evaluate the rate $\lambda$ by using the 
nonlinear formula
\be
\label{eq:lambda_known}
\lambda=\sqrt{x^2+y^2}\,.
\ee
It is clear that this procedure is experimentally much less 
demanding than the previous alternative.

Let us change a bit our perspective and interpret the same setup 
as an experiment in which two copies of the channel are used 
in each run of the experiment. In other words, let us assume that the test 
state is a two-qubit state 
$\ket{+}\bra{+}\otimes\ket{+}\bra{+}$ entering the channel 
$\cE_\lambda\otimes\cE_\lambda$. The two-output state is then measured
in a factorized measurement $\sigma_x\otimes\sigma_y$. The general
process measurement scheme of such two-copies type leads  to a statistics 
\be
p(F)=\tr{[(\Omega_\lambda\otimes\Omega_\lambda)F]}\,,
\ee
where $F$ are the operators (elements of the PPOVM acting on a four-qubit system) 
describing the individual observations. In the considered case the outcomes
are associated with the operators
$$F_{\pm,\pm}:=\ket{++}\bra{++}\otimes \ket{\pm x,\pm y}\bra{\pm x,\pm y}\,,$$
where $\ket{\pm x},\ket{\pm y}$ are the eigenvectors of $\sigma_x$ and $\sigma_y$,
respectively. In particular,
\be
p(F_{\pm,\pm})=\frac{1}{4}[1\pm {\rm Re}(\omega_{01})]\cdot[1\pm {\rm Im}(\omega_{01})]\,,
\ee
and hence the observed probabilities are nonlinear (quadratic) functions
of the channel's parameters $\omega_{01}$ and $\omega_{10}$. Taking into
account the fact that $x={\rm Re}(\omega_{01})$ and 
$y={\rm Im}(\omega_{01})$, we see how to play with the observed probabilities
to recover Eq. (\ref{eq:lambda_known}). 

The message of this example is that certain nonlinear parameters can be accessed
in experiments in which two copies of the channel are tested simultaneously.
The question that remains is whether  the value of $\lambda$ can also be
observed directly by measuring a single experimental quantity $Q$ only, i.e.,
whether there exists a single observable $Q$ such that $\lambda=f(\tr{[(\Omega_\lambda\otimes\Omega_\lambda) Q]})$.

To this end let us consider yet another experiment in which instead of initializing the
two-qubit test system in the state $\ket{++}\bra{++}$, we send the entangled vector state $\ket{\phi_+}$
(with the definition $\ket{\phi_{\pm}}:=(\ket{01}\pm\ket{10})/\sqrt{2}$) through
the channel $\cE_\lambda\otimes\cE_\lambda$. 
The output state reads
\be
\nonumber
\omega^\prime=\frac{1}{2}[\ket{01}\bra{01}+\ket{10}\bra{10}+
|\omega_{01}|^2(\ket{01}\bra{10}+\ket{10}\bra{01})]\,,
\ee
where $(\ket{01}\bra{10}+\ket{10}\bra{01})/2=
\ket{\phi_+}\bra{\phi_+}-\ket{\phi_-}\bra{\phi_-}=:Z$ is a Hermitian 
operator and $\tr{Z(\ket{01}\bra{01}+\ket{10}\bra{10})}=0$. That is,
$Z$ has the expectation value
\be
\tr{[Z\omega^\prime]}=\frac{1}{2}|\omega_{01}|^2\,,
\ee
and hence the statistics of the nonlocal, two-valued measurement 
associated with the Hermitian operator $Z$ uniquely and directly
determines the parameter $\lambda=\sqrt{2\l Z\r}$. This is exactly 
the direct method for estimating the inverse decoherence rate that we were looking for.
This measurement determines only a single independent parameter which can be immediately and uniquely
transformed into the value of $\lambda$.

Let us note that the process POVM considered above is described by
$F_\pm=\ket{\phi_+}\bra{\phi_+}\otimes\ket{\phi_\pm}\bra{\phi_\pm}$ and
$F_o=\ket{\phi_+}\bra{\phi_+}\otimes (\ket{00}\bra{00}+\ket{11}\bra{11})$
being the output with vanishing probability of appearance. It is 
straightforward to verify that
\be
\tr{[(\Omega_\lambda\otimes\Omega_\lambda)Q]}=\frac{1}{2}|\omega_{01}|^2\,,
\ee
where $Q=F_+-F_-=\ket{\phi_+}\bra{\phi_+}\otimes Z$. In conclusion, given that
the decoherence basis is known, the value of $\lambda$ can be directly
accessed and identified in a single, effectively two-valued, measurement.
%%%%%%%%%%%%%%%%%%%%%%%%%%%%%%%%%%%%%%%%%%%%%%%%%%%%%%%%%%%%%%%%%%%%%%%%%%%%%
\subsection{Unknown decoherence basis}
%%%%%%%%%%%%%%%%%%%%%%%%%%%%%%%%%%%%%%%%%%%%%%%%%%%%%%%%%%%%%%%%%%%%%%%%%%%%%
In this section we will ask the same question about the inverse decoherence rate but
relax the assumption that the decoherence basis is known. Since
the previous problem is a special case of this one, it is immediately evident that there can be no 
single-copy experiment which  measures the value of $\lambda$ directly. 
Therefore, let us focus on the existence of direct two-copy experiments.
Our goal is to identify the observable $Q$ that one needs to measure in order to
access the value of $\lambda$ directly. 

Let us start with a setup in which no ancilla is involved, i.e.,
the test state $\varrho\in\cS(\cH\otimes\cH)$ passes the
channel $\cE_\lambda\otimes\cE_\lambda$, and we 
measure the expected value of a Hermitian operator $M$ 
defined on $\cH\otimes\cH$. Clearly, the scheme should be
basis-independent, meaning that the operators $\varrho$ and $M$
must  both be $U\otimes U$-invariant (have the same form 
in any basis of $\cH$). This implies that
$\omega=q \varrho_++(1-q)\varrho_-$, where $\varrho_\pm:=P_\pm/d_\pm$,
$P_\pm$ are projections onto symmetric and antisymmetric subspaces of 
$\cH\otimes\cH$, respectively, and $d_\pm:=d(d\pm 1)/2$ are the dimensions
of the subspaces spanned by $P_\pm$. This is the well-known family
of Werner states. Let us stress that in this section
we assume $d=2$, i.e., $P_+=\ket{00}\bra{00}+\ket{11}\bra{11}
+\ket{\phi_+}\bra{\phi_+}$ and $P_-=\varrho_-=\ket{\phi_-}\bra{\phi_-}$.
Similarly, any Hermitian operator $M=aI+b S$ (with $a,b$ real), 
where $S:=P_+-P_-$ is the swap operator, is $U\otimes U$ invariant.

Let us recall once more that the expressions for $P_\pm$ have
the same form in any basis of $\cH$. Therefore,
\be
(\cE_\lambda\otimes\cE_\lambda)[P_+]&=&A+B+\lambda^2 Z\,,\\
(\cE_\lambda\otimes\cE_\lambda)[P_-]&=&B-\lambda^2 Z\,,
\ee
where we used 
\be
A&:=&\ket{00}\bra{00}+\ket{11}\bra{11}\,,\\
B&:=&\frac{1}{2}(\ket{01}\bra{01}+\ket{10}\bra{10})\,,
\ee 
and the Hermitian operator 
$Z=(\ket{01}\bra{10}+\ket{10}\bra{01})/2$ is defined as
in the previous section. Since $S=A+Z=\ket{00}\bra{00}+\ket{11}\bra{11}+
\ket{\phi_+}\bra{\phi_+}-\ket{\phi_-}\bra{\phi_-}$, we find that
$\tr{[SA]}=2$, $\tr{[SZ]}=1$ and $\tr{[SB]}=0$. Moreover, $\tr{A}=2$, 
$\tr{B}=1$ and $\tr{Z}=0$. Using all these identities we obtain
\be
\nonumber
\l M\r &=&\tr{\left[(\cE_\lambda\otimes\cE_\lambda)[\varrho]M\right]}
\\ \nonumber
&=&\frac{q}{3}\tr{[M(A+B+\lambda^2 Z)]}+(1-q)\tr{[M(B-\lambda^2 Z)]}
\\ \nonumber &=&\frac{q}{3}(2a+2b+a+b\lambda^2)+(1-q)(a-b\lambda^2)
\\ &=&a+\frac{1}{3}[2q+\lambda^2(4q-3)]b\,.
\ee
Setting $a=0$, $b=1$ (i.e., $M=S$) we obtain the direct estimation formula
for the inverse decoherence rate in its simplest form
\be
\lambda=\sqrt{\frac{3\l S\r-2q}{4q-3}}\,.
\ee

\begin{figure}[htbp]
\centering

    \includegraphics[scale=0.3]{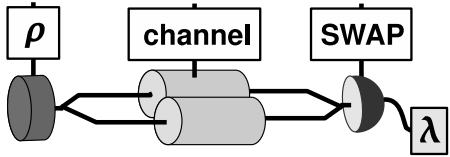}
\caption[]{Direct inverse decoherence rate measurement, consisting of a preparation of a Werner state, two applications of the decoherence channel, and a SWAP measurement.}
\label{fig:experiment}
\end{figure}

Let us stress that the basis-dependent operators $A,B,Z$ need not be measured;
only the  basis-independent swap operator $S$ has to be measured.
The whole
experiment, depicted in Fig. \ref{fig:experiment}, is indeed basis-independent and directly determines 
the value of the nonlinear parameter $\lambda$, as desired. 

%%%%%%%%%%%%%%%%%%%%%%%%%%%%%%%%%%%%%%%%%%%%%%%%%%%%%%%%%%%%%%%%%%%%%%%%%%%%%
\section{Average inverse decoherence rate for qudits}
%%%%%%%%%%%%%%%%%%%%%%%%%%%%%%%%%%%%%%%%%%%%%%%%%%%%%%%%%%%%%%%%%%%%%%%%%%%%%
The case of $d$-dimensional quantum systems (qudits) is a lot more
complicated because each pair of decoherence basis elements
$\ket{j},\ket{k}$ is associated with its own inverse decoherence
rate $\lambda_{jk}=|\omega_{jk}|$. Let us perform the same experiment
as in the qubit case (see  Fig. \ref{fig:experiment}). Suppose the initial test state is 
\be
\label{eq:werner_state}
\varrho=q\varrho_++(1-q)\varrho_-\,,
\ee
where, as before, $\varrho_\pm=P_\pm/d_\pm$. The procedure reads as follows: Apply the channel 
$\cE_{\lambda}\otimes\cE_{\lambda}$, where ${\lambda}$ now 
denotes all the inverse decoherence rates $\lambda_{jk}$. And then measure
the expectation value of the swap operator $S$.

Let us define for $j\neq k$ the following (mutually orthogonal) unit vectors
\be
\phi_{\pm,jk}:=\frac{1}{\sqrt{2}}(\ket{jk}\pm\ket{kj})\,.
\ee
Then
\be
P_+&=&\sum_j \ket{jj}\bra{jj}+\sum_{j<k}\ket{\phi_{+,jk}}\bra{\phi_+,jk}\,,\\
P_-&=&\sum_{j<k}\ket{\phi_{-,jk}}\bra{\phi_{-,jk}}\,.
\ee
Since $\cE_{\lambda}[\ket{j}\bra{k}]=\omega_{jk}\ket{j}\bra{k}$
and $\omega_{jj}=1$, we have that
\be
(\cE_\lambda\otimes\cE_\lambda)[P_+]&=&
\sum_j\ket{jj}\bra{jj}+\\
\nonumber & & +
\frac{1}{2}\sum_{j\neq k}\left(\ket{jk}\bra{jk}+\lambda_{jk}^2\ket{jk}\bra{kj}\right)\,,\\
(\cE_\lambda\otimes\cE_\lambda)[P_-]&=&
\frac{1}{2}\sum_{j\neq k}\left(\ket{jk}\bra{jk}-\lambda_{jk}^2\ket{jk}\bra{kj}\right)\,.
\ee
Let us note that $|\omega_{jk}|=\lambda_{jk}=\lambda_{kj}=|\omega_{kj}|$ and hence
\be
\tr\left(S(\cE_\lambda\otimes\cE_\lambda)[P_+]\right)&=&d+0+
\sum_{j<k}\lambda_{jk}^2\,,\\
\tr\left(S(\cE_\lambda\otimes\cE_\lambda)[P_-]\right)&=&0-
\sum_{j<k}\lambda_{jk}^2\,.
\ee
Now it is straightforward to see that
the measured average value of the swap operator 
in the considered experiment equals 
\be
\nonumber
\l S\r&=&\tr{\left(S(\cE_\lambda\otimes\cE_\lambda)[\varrho]\right)}\\
&=& \frac{q}{d_+}(d+\sum_{j<k}\lambda_{jk}^2)-
\frac{1-q}{d_-}\sum_{j<k}\lambda_{jk}^2\,.
\ee
Since $d_-=d(d-1)/2$ coincides with the number of pairs $jk$ we may
define the average decoherence rate 
$\overline{\lambda^2}:=\frac{1}{d_-}\sum_{j<k}\lambda_{jk}^2$ and obtain
\be
\l S\r=[\overline{\lambda^2}(qd^2-d_+)+qd]/d_+\,;
\ee
so the direct estimation formula for
$\overline{\lambda^2}$ reads
\be
\label{eq:estimation_formula}
\overline{\lambda^2}=\frac{d_+\l S\r-qd}{qd^2-d_+}
=\frac{(d+1)\l S\r - 2q}{2qd-(d+1)}\,.
\ee

%%%%%%%%%%%%%%%%%%%%%%%%%%%%%%%%%%%%%%%%%%%%%%%%%%%%%%%%%%%%%%%%%%%%%%%%%%%%%
\section{Double-commutator master equation}
%%%%%%%%%%%%%%%%%%%%%%%%%%%%%%%%%%%%%%%%%%%%%%%%%%%%%%%%%%%%%%%%%%%%%%%%%%%%%
In case of qubits it was shown {\cite{ziman2005}} that master equations
inducing pure decoherence channels are of the double-commutator form
\begin{eqnarray}
 \frac{d}{dt} \varrho = \cL (\varrho) = - \frac{i}{\hbar} [\varrho, H] -
 \frac{1}{2 \hbar^2 \gamma} [[\varrho, H], H] \hspace{0.25em}, 
 \label{eq:LimbladianDoubleComm}
 \label{eq:AvgDecoRateQudit}
\end{eqnarray}
where $H$ is a Hermitian operator.
Let us call this operator ``Hamiltonian'' and
its eigenvalues ``eigenenergies''. The channels $\cE = e^{\cL t}$ are pure
decoherence channels, and the eigenbasis of $H$ represents the decoherence basis.
This master equation has perfect sense in any dimension (c.f. \cite{OpenQS,PhysRevA.44.5401}), and it is natural to
ask whether our previous analysis can help somehow to directly measure the
parameter $\gamma$.

Suppose $h_j$ are the eigenenergies of $H$,
and denote their differences by $\Delta_{jk}:=h_j-h_k$.
Then the master equation (\ref{eq:LimbladianDoubleComm}) reads
\begin{equation}
 \label{eq:LimbladianDoubleCommExplicit} \mathcal{L} \left( \rho \right) =
 \sum^d_{j,k = 0} \left( 1 - \delta_{jk} \right)  \frac{\Delta_{jk}}{\hbar} \left( i - \frac{\Delta_{jk}}{2 \hbar
 \gamma} \right) \rho_{jk}  \ket{j} \bra{k} .
\end{equation}
Setting $\omega_{jk}=\lambda_{jk}e^{i\varphi_{jk}}$ 
in Eq.~(\ref{eq:DensMatTransformation}), it follows 
from $\cE=e^{\cG}$ that the generator of a general pure decoherence
process has the form
\begin{equation}
 \label{eq:LimbladianPureDeco} \mathcal{G} \left( \rho \right) = \sum^d_{j,k
 = 0} \left( 1 - \delta_{jk} \right)  \left( \ln \lambda_{jk} 
+ i \varphi_{jk} \right) \rho_{jk}
 \ket{j} \bra{k} .
\end{equation}
Comparing the real and imaginary parts of
Eqs.~(\ref{eq:LimbladianDoubleCommExplicit}) and (\ref{eq:LimbladianPureDeco})
we have
\begin{equation}
 \label{eq:DoubleCommuttator-phi} \varphi_{jk} = \Delta_{jk} / \hbar
\end{equation}
and
\begin{equation}
 \label{eq:DoubleCommuttator-gamma} \lambda_{jk} = \exp \left(
 \frac{- \Delta^2_{jk}}{2 \hbar^2 \gamma} \right) .
\end{equation}
With the help of Eq.~(\ref{eq:DoubleCommuttator-gamma}) one can express the average
inverse decoherence rate as a function of $\gamma$,
\begin{equation}
 \label{eq:Eq4gamma} \overline{\lambda^2} = \frac{1}{d_-} \sum_{j < k} \exp
 \left( - \frac{\Delta_{j k}^2}{\hbar^2 \gamma} \right).
\end{equation}

The question is whether it is possible to invert this
formula. If this can be done then the experimental
observation of $\overline{\lambda^2}$ via measurement of $\l S\r$
(as described in the previous section) directly determines  
$\gamma=f(\overline{\lambda^2})$. Clearly, 
if complete information about the  
differences $\Delta_{jk}$ of the eigenenergies of $H$ is available (with the eigenbasis
remaining unspecified), Eq. (\ref{eq:Eq4gamma}) specifies the parameter $\gamma$ implicitly.
Unfortunately, we are not able to explicitly perform
the inversion, nor to specify the conditions under which
such an inversion is possible, for an arbitrary Hamiltonian. Let us note
that once the parameter $\gamma$ is specified, the knowledge of
$\Delta_{jk}$ would allow us to determine all inverse decoherence rates 
$\lambda_{jk}$. 

A special case in which the inversion is in fact possible is a Hamiltonian which is equally gapped 
(for instance, that of a linear harmonic oscillator), for which
\begin{eqnarray}
 \Delta_{jk} =  (j - k) \Delta\,,\label{eq:deltas}
\end{eqnarray}
where $\Delta$ is the elementary energy gap. The absolute values of the differences
$|j-k|$ take values $n=1,\dots,d-1$, and each of these appears exactly $N_n=d-n$ times 
in the sum present in Eq.~(\ref{eq:Eq4gamma}). Therefore, 
Eq.~(\ref{eq:Eq4gamma}) can be rewritten in the form
\begin{equation}
 \label{eq:Eq4xEquiGapped} \overline{\lambda^2} = \frac{1}{d_-}  \sum_{n
 = 1}^{d - 1} \left( d - n \right) K^{n^2} \,,
\end{equation}
where $K:=\exp{\left[-\Delta^2/(\hbar^2\gamma)\right]}$. The direct 
estimation formula for $gamma$ reads
\begin{equation}
 \label{eq:gamma} \gamma = \frac{- \Delta^2}{\hbar^2 \ln  K_r }\,,
\end{equation}
where 
\begin{equation}
 \label{eq:Rk} K_r \assign 
{\rm Root}_r
 \left[ \overline{\lambda^2} d_- - \sum^{d - 1}_{n = 1} \left( d -
 n \right) K^{n^2} \right],
\end{equation}
is the $r$th root of the above polynomial in the variable $K$  with $r=1$ for even Hilbert space dimension $d$ and $r=2$ for odd $d$. The ordering of roots is such that real roots precede complex ones, with the real roots ordered by size. We have verified numerically for all $d$ up to 32, that the $r$th root, Eq. (\ref{eq:Rk}), is the unique positive real solution of Eq. (\ref{eq:Eq4xEquiGapped}). 

Morevoer, the identity $K_r=\exp{\left[-\Delta^2/(\hbar^2\gamma)\right]}$
and Eqs. (\ref{eq:DoubleCommuttator-gamma}) and (\ref{eq:deltas}) enable
us to write down the estimation formula for the individual off-diagonal elements
\begin{eqnarray}
 \lambda_{jk} = K_r^{|j-k|^2/2}\,.
\end{eqnarray}
Let us note that in this case the knowledge of the actual value of the gap $\Delta$ is not necessary.

In summary, the method of direct observation of the average inverse decoherence
rate $\overline{\lambda^2}$ can be employed [cf. Eq.~(\ref{eq:Eq4gamma})] 
to estimate the decoherence decay parameter $\gamma$ which occurs 
in the double-commutator master equation given in 
Eq.~(\ref{eq:LimbladianDoubleComm}), provided that the additional
knowledge on the structure of energy gaps (spectrum) is available. 
This information can then also be used to determine directly all individual inverse 
decoherence rates $\lambda_{jk}$.

%%%%%%%%%%%%%%%%%%%%%%%%%%%%%%%%%%%%%%%%%%%%%%%%%%%%%%%%%%%%%%%%%%%%%%%%%%%%%
\section{Discussion and conclusions}
%%%%%%%%%%%%%%%%%%%%%%%%%%%%%%%%%%%%%%%%%%%%%%%%%%%%%%%%%%%%%%%%%%%%%%%%%%%%%
In this paper we proposed an experimental setup that can directly determine the
value of the average inverse decoherence rate, provided that
the channel under consideration is a so-called pure decoherence channel.
The experiment consists of the following steps: 
\begin{enumerate}
\item Prepare a bipartite Werner state with $q\neq d_+/d^2$ 
[see Eq.(\ref{eq:werner_state})].
\item Apply the channel $\cE_\lambda\otimes\cE_\lambda$.
\item Perform the SWAP measurement and record the expectation value.
\item Apply the estimation formula (\ref{eq:estimation_formula}).
\end{enumerate}
Let us recall that the SWAP measurement is a two-valued measurement,
and hence by definition, it can fix only one parameter of $\cE_\lambda$.
Therefore, no other information about decoherence beyond its average rate
is revealed. For instance, the decoherence basis remains 
as unknown as it was at the beginning. 

\begin{figure}[htbp]
\centering
   \includegraphics[scale=0.3]{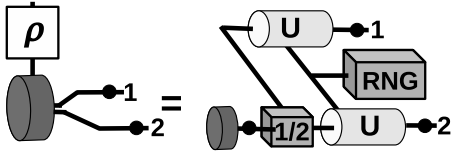}
   \caption[]{Preparation of a two-partite Werner state using a single-particle source and sampling 
of a random unitary channel.}
\label{fig:preparation}
\end{figure}

Let us stress that no complicated system source is needed in order to
accomplish this measurement. In fact, any state $\varrho$ can be turned
into a Werner state by a so-called twirling transformation 
\be
\omega_q=\int dU U\otimes U \varrho U^\dagger\otimes U^\dagger\,.
\ee
Moreover, instead of a source of bipartite systems
we can employ two copies $\eta\otimes\eta$ produced by the same single particle
source. Applying the same random unitary channel on each of the copies 
will result in a Werner state with some initial value of $q$ (see Fig. \ref{fig:preparation}).

\noindent This
value is related to the purity of the single particle source by
\be
q=\frac{1}{2}(1+\tr{\eta^2})\,.
\ee
This is the only information about the test state that we really need,
and it is well known how to measure the purity directly \cite{brun_measuringNLfunctions, shalm_measuringPurity}:
Take two copies, perform the SWAP measurement and relate $\tr{\eta^2}=\tr{S\eta\otimes\eta}=\l S\r$. If one uses this way of preparing the Werner state, the accessible values of $q$ are 
limited by $d_{+} /d^2 \le q\le 1$. Furthermore, let us note that the mixing procedure 
presupposes that no information on the actual sequence of the random $U$s is used in the estimation. Otherwise,
the procedure could not be seen as a preparation of a Werner state.

The key operation in the preparation procedure is the twirling.
It is experimentally difficult to really sample from the Haar measure
over unitary channels. However, this is not really needed, as
the integration can be replaced by an average over only a
discrete number of unitary channels. This construction is known
as a unitary $k$-design  \cite{k-design,k-design2,k-design3}. In our particular case we are interested
in a so-called 2-design, which is any probability distribution
$\{p_1,\dots,p_n\}$ distributed over unitary operators 
$\{U_1,\dots,U_n\}$ such that for all operators $X$,
\be
\nonumber
\int dU (U^{\otimes 2}) X (U^{\otimes 2})^\dagger=
\sum_{j\in J} p_j (U_j\otimes U_j) X
(U_j^\dagger\otimes U_j^\dagger)\, .
\ee
In particular \cite{k-design}, for a $d$-dimensional quantum system 
the uniform 2-design (i.e. $p_j=1/n$) consists of $n\geq d^4-2d^2+2$ 
unitaries.

Let us summarize which resources are sufficient for direct 
observation of the inverse decoherence rate: The experimenter needs 
a single particle source of (not completely mixed) states; he 
must be able to implement a finite number of (undisclosed) unitaries forming
the support of a 2-design; and finally, he must be able to perform
the SWAP measurement.

In conclusion, we have addressed the problem of direct experimental
measurement of a specific nonlinear channel parameter, the average inverse decoherence
rate. We have shown that experiments which collect statistics on
two copies of the channel perfectly serve this purpose. Since many interesting
channel properties 
(quantum capacities, rates, etc.) are not linear
and cannot be accessed directly in single-copy experiments, it is
of practical importance to understand more deeply the theory of direct
estimation of nonlinear channel parameters. The present paper represents 
an important case study in this regard.

%%%%%%%%%%%%%%%%%%%%%%%%%%%%%%%%%%%%%%%%%%%%%%%%%%%%%%%%%%%%%%%%%%%%%%%%%%%%%
%%%%%%%%%%%%%%%%%%%%%%%%%%%%%%%%%%%%%%%%%%%%%%%%%%%%%%%%%%%%%%%%%%%%%%%%%%%%%
%%%%%%%%%%%%%%%%%%%%%%%%%%%%%%%%%%%%%%%%%%%%%%%%%%%%%%%%%%%%%%%%%%%%%%%%%%%%%
%%%%%%%%%%%%%%%%%%%%%%%%%%%%%%%%%%%%%%%%%%%%%%%%%%%%%%%%%%%%%%%%%%%%%%%%%%%%%

%%%%%%%%%%%%%%%%%%%%%%%%%%%%%%%%%%%%%%%%%%%%%%%% ACKNOWLEDGMENT %%%%%%%%%%%%
\section{Acknowledgements} 
This work was supported by EU integrated project Q-ESSENCE, 
VEGA 2/0092/11 (TEQUDE) and APVV-0646-10 (COQI). 
M.Z. acknowledges a support of the SCIEX Fellowship 10.271. P.R. acknowledges a support of the \v{S}tefan Schwarz fellowship.
%%%%%%%%%%%%%%%%%%%%%%%%%%%%%%%%%%%%%%%%%%%%%% BIBLIOGRAPHY %%%%%%%%%%%%%%%%
%\section*{References}

%%%%%%%%%%%%%%%%%%%%%%%%%%%%%%%%%%%%%%%%%%%%%%%%%%%%%%%%%%%%%%%%%%%%%%%%%%%%%

\end{document}